\documentclass[conference]{IEEEtran}
\IEEEoverridecommandlockouts
\usepackage{cite}
\usepackage{graphicx}
\usepackage{subcaption}

\usepackage{amsmath,amssymb,amsfonts}
\usepackage{algorithmic}
\usepackage{graphicx}
\usepackage{textcomp}
\usepackage{xcolor}

\usepackage[margin=0.780in]{geometry}

\def\BibTeX{{\rm B\kern-.05em{\sc i\kern-.025em b}\kern-.08em
    T\kern-.1667em\lower.7ex\hbox{E}\kern-.125emX}}
\begin{document}

\title{Optimal Placement of Transmissive RIS in the Near Field for Capacity Maximization in THz Communications}
\author{\IEEEauthorblockN{Nithish Sharvirala, Amine Mezghani, \textit{Member, IEEE}, and Ekram Hossain, \textit{Fellow, IEEE}} 
\thanks{N. Sharvirala is with Indian Institute of Technology Patna (email: sharvirala\_2001ee65@iitp.ac.in). 
This work was done during his Mitacs Globalink Research Internship at the University of Manitoba, Canada. 
A. Mezghani and E. Hossain are with the Department of Electrical and Computer Engineering at the University of Manitoba, Canada (emails: \{Amine.Mezghani, Ekram.Hossain\}@umanitoba.ca).}

}

\maketitle

\begin{abstract}
This study centers on Line-of-Sight (LoS) MIMO communication enabled by a Transmissive Reconfigurable Intelligent Surface (RIS) operating in the Terahertz (THz) frequency bands. The study demonstrates that the introduction of RIS can render the curvature of the wavefront apparent over the transmit and receive arrays, even when they are positioned in the far field from each other. This phenomenon contributes to an enhancement in spatial multiplexing. Notably, simulation results underline that the optimal placement of the RIS in the near-field is not solely contingent on proximity to the transmitter (Tx) or receiver (Rx) but relies on the inter-antenna spacing of the Tx and Rx.

\end{abstract}



\section{Introduction}
Recent shifts in the patterns of multimedia consumption have caused a substantial surge in mobile data traffic. The growing popularity of wireless networks has led to a greater need for high-speed wireless communication systems [1]. To accommodate the elevated data rates anticipated in future wireless communication systems, it becomes essential to explore advanced solutions at the physical layer and to seek out new spectral bands. Among the spectrum bands showing promise for facilitating ultra-high-speed communication is the Terahertz (THz) band, stretching from 0.06 to 10 THz. The THz band is characterized by its extensive bandwidth, which varies from tens of gigahertz to several terahertz, in accordance with the transmission distance [2].

Nonetheless, operating in this frequency band does pose certain challenges in regard to radio signal propagation. Moreover, its possible applications are largely constrained to shorter-range communications. Successful communication in this band is heavily dependent on maintaining a direct line-of-sight (LoS) connection or leveraging robust signal reflections. However, due to the combined characteristics of a relatively small wavelength and restricted transmission range, this opens up an opportunity for employing multiple-input multiple-output (MIMO) technology with practical physical dimension arrays in LoS scenarios. In fact, when the curvature of the radio wavefront becomes apparent across the transmit and receive arrays, the channel matrix's rank improves, even under pure LoS conditions. Therefore, LoS conditions are well-suited for THz communication, offering advantages such as minimal path loss and improved spatial multiplexing [3].

 THz communications have a significant role to play in wireless backhaul networks. They can serve as an alternative to fiber cables or complement existing fiber infrastructure to enhance wireless backhaul. In cases where replacing fiber is considered, it's essential to establish a direct LoS connection with the maximum possible distance [4]. Consequently, it may not always be feasible to maintain both base stations within each other's near-field region. This could potentially result in the loss of the beneficial exploitation of wavefront curvature to amplify the rank of the channel matrix.

Reconfigurable Intelligent Surfaces (RISs) have surfaced as an innovative type of passive, non-regenerative relay technology. It can be programmed to reflect or refract electromagnetic waves with desired phase shift [5]. Operating in THz bands, an RIS generates a radiating near-field, known as the Fresnel zone, that extends over several meters. Placing an adequately sized RIS between base stations has the potential to make the wavefront curvature discernible across the arrays, leading to an enhancement in the channel matrix's rank. 

Papers [6,7] establish that the Intelligent reflecting surface (IRS)-assisted system's capacity is optimum only when the IRS is in the vicinity of the transmitter (Tx) or receiver (Rx). However, both these works assume the Tx and Rx in the far field of the IRS, where the wavefront is approximated as planar and does not influence the rank of the channel matrix. As a result, the optimal placement of the RIS in the near field might not align with its optimal position in the far field, which serves as a driving motivation for this present study.

This work considers a transmissive RIS-assisted MIMO system operating in THz bands, with LoS Tx-RIS and RIS-Rx connections, and without a direct path between the Tx and Rx due to obstruction by the RIS. The key idea of the paper is that since the degrees of freedom (DoF) in LoS MIMO escalate as the wavefront curvature becomes noticeable over the transmit and receive arrays, concatenating two LoS MIMO links via a RIS—which acts as the receiver for the first link and the transmitter for the second link—ensures both links in its Fresnel zone, thereby making the wavefront curvature noticeable over the Tx and Rx even though they are in the far-field to each other.

The main contributions of this work include:
\begin{itemize}
  \item The application of RIS to extend the near-field, thereby enhancing the DoF available for LoS spatial multiplexing.
  \item The computation of the optimal positioning of the RIS in the near-field.
  \item The examination of variation of optimal position with inter-antenna distance using the singular values of the effective channel matrix.
\end{itemize}

\section{System Model and Assumptions}

\begin{figure}
\centering
\includegraphics[width=0.9\linewidth]{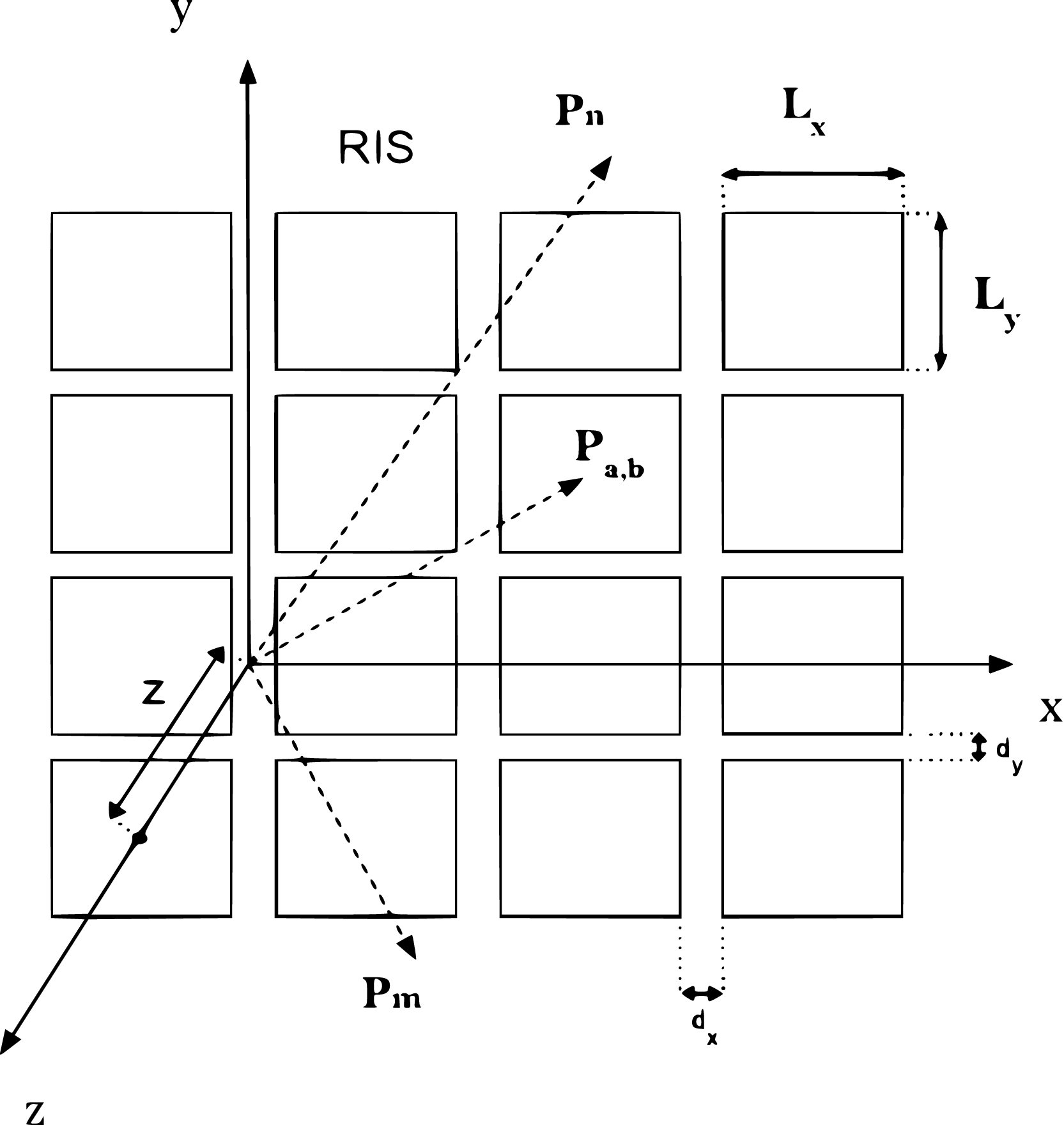}
\caption{Depiction of the RIS geometry assumed in the system model.}
\label{fig:my_picture}
\end{figure}

\subsection{Signal Model}

Consider a THz transmissive  RIS system equipped with $N_t\geq1$ antennas at the Tx and $N_r\geq1$ antennas at the Rx. The RIS is positioned in the $xy$-plane and comprises a total of $W = W_x\times W_y$ elements. Every element possesses dimensions of $L_x$ in width and $L_y$ in length. The gaps between neighboring elements are denoted as $d_x$ along the x-axis and $d_y$ along the y-axis.  
Assuming the RIS to be lossless, the transmitting coefficient of the $(a, b)$th element is represented as $\beta_w=e^{j\varphi_{a,b}}$, where $\varphi_{a,b} \in [0, 2\pi)$.

The channel matrix for the direct link that connects the Tx and the Rx is denoted by $\boldsymbol{H} \in \mathbb{C}^{N_r \times N_t}$.
Similarly, $\boldsymbol{T} \in \mathbb{C}^{W \times N_t}$ signifies the channel matrix from the Tx to the RIS, $\boldsymbol{R} \in \mathbb{C}^{N_r \times W}$ the channel matrix from the RIS to the Rx, ${\boldsymbol{\it \Phi}} \in \mathbb{C}^{W \times W} $ the diagonal transmission matrix of the RIS, where 
 ${\boldsymbol{\it \Phi}} = {diag}\{\beta_1, ..., \beta_W\}$. Hence, the effective MIMO channel matrix for the system setup with only the Tx-RIS-Rx link can be represented as $\boldsymbol{\tilde{H} =  R{\boldsymbol{\it \Phi}} T}$. 

Let us denote the transmitted signal vector as $\boldsymbol{s} \in \mathbb{C}^{N_t \times 1}$. Define the transmit signal covariance matrix as $\boldsymbol{R_s} \triangleq \mathbb{E}[\boldsymbol{s}\boldsymbol{s}^H] \in \mathbb{C}^{N_t \times N_t}$, with $\boldsymbol{R_s} \succeq 0$, accounting for its positive semi-definite nature. A Tx with an average sum power constraint, which can be expressed as $\mathbb{E}[\|\boldsymbol{s}\|^2] \leq P$, equivalently represented as $\text{tr}(\boldsymbol{R_s}) \leq P$ is considered. The received signal vector, represented as $\boldsymbol{y} \in \mathbb{C}^{N_r \times 1}$, can be described as
\begin{equation}
    \boldsymbol{y} = \boldsymbol{\tilde{H}s} + \boldsymbol{n} = \boldsymbol{(R{\boldsymbol{\it \Phi}} T)s} + \boldsymbol{n}
\end{equation}where $\boldsymbol{n} \sim \mathcal{CN}(0, \sigma^2 \boldsymbol{I}_{N_r})$ characterizes the additive noise at the receiver, $\sigma^2$ represents the average noise power.

\subsection{Channel Model}

We assume that both the Tx and Rx are in the Fresnel zone of the RIS given by $0.62\sqrt{L^3_{\text{RIS}}/\lambda} < D \leq 2L^2_{\text{RIS}}/\lambda$ [8], where D is the distance measured from the RIS and 
$L_{\text{RIS}} = \max(W_x L_x + (W_x - 1) {d}_x, W_y L_y + (W_y - 1) {d}_y)$ is the maximum physical dimension of the RIS. We model the channel considering the spherical wavefront of the radiated waves.

Antenna arrays are conventionally represented as an ensemble of point radiators, whereas an RIS consists of rectangular elements with substantial dimensions. Consider the RIS with the center of the $(0, 0) {\text{th}}$ RIS element placed on the z-axis of the coordinate system, as shown in Fig. 1. Within $(a,b)$th RIS element, the transmitting coefficient $e^{j\varphi_{a,b}}$ remains constant.
Hence, the position vector of the $(a, b)$th RIS element is $\mathbf{p}_{a,b}$ = $(a\overline{d}_x,b\overline{d}_y,z)$, where $\overline{d}_x = d_x + L_x$, $\overline{d}_y = d_y + L_y$ and $z$ is the z-coordinate corresponding to the centre of the $(0, 0) {\text{th}}$ RIS element. The locations of arbitrary transmit and receive antennas are $(D_m, \theta_m, \phi_m)$ and $(D_n, \theta_n,\phi_n)$, 
respectively. Therefore, their Cartesian coordinates are 

\begin{equation}
\begin{aligned}
\mathbf{p}_{m} &= (D_{m} \cos \phi_{m} \sin \theta_{m}, D_{m} \sin \phi_{m} \sin \theta_{m}, \\
&\qquad D_{m} \cos \theta_{m}),
\hspace{1.55cm} m = 1, \ldots, N_t
\end{aligned}
\end{equation}

\begin{equation}
\begin{aligned}
\mathbf{p}_{n} &= (D_{n} \cos \phi_{n} \sin \theta_{n},  D_{n} \sin \phi_{n} \sin \theta_{n}, \\
&\qquad D_{n} \cos \theta_{n}),
\hspace{1.45cm} n = 1, \ldots, N_r,
\end{aligned}
\end{equation}
where $D_{m}$ and $D_{n}$ represent the distances from the origin to the $m$ th transmit and $n$ th receive antennas, the azimuth and polar angles are represented by $\phi$ and $\theta$. The entries of the channel from the Tx to the RIS are denoted as [8]

\begin{equation}
[\boldsymbol{T}]_{w,m} = \sqrt{\mathbf{PL}_{a,b}^{m}} e^{-jkD_{a,b}^{m}}.
\end{equation}
In (4), $k = 2\pi/\lambda$ denotes the wavenumber, where $\lambda$ signifies the carrier wavelength, $D_{a,b}^{m} \triangleq \left\|\mathbf{p}_{m} -  \mathbf{p}_{a,b}\right\|$ represent the distance between the $m$ th transmit antenna and the $(a,b)$th RIS element, while $\mathbf{PL}_{a,b}^{m}$ = $(G_{m}L_xL_ye^{-\kappa_{\text{abs}}(f)(D_{a,b}^{m})})/4\pi (D_{a,b}^{m})^2$ is the respective path loss with $G_{m}$ and $\kappa_{\text{abs}}(f)$ denoting the \\$m$ th transmit antenna's gain and the molecular absorption coefficient at the carrier frequency $f$.
The entries of $\boldsymbol{R} \in \mathbb{C}^{N_r \times W}$ can be expressed as [9, eq. 7]

\begin{equation}
[\boldsymbol{R}]_{n,w} = \sqrt{G_{n}e^{-\kappa_{\text{abs}}(f)(D_{a,b}^{n})}}
\frac{L_xL_y}{j\lambda D_{a,b}^{n}}F(\theta_w^{n}) e^{jkD_{a,b}^{n}},
\end{equation}
where $j$ is the imaginary unit, ${G_{n}}$ denotes the $n$ th receive antenna's gain, while $D_{a,b}^{n} \triangleq \left\|\mathbf{p}_{n} -\mathbf{p}_{a,b}\right\|$, $F(\theta_w^{n}) = {(1+\cos{\theta_w^{n}})}/{2}$ is the leaning factor according to the Fresnal-Kirchhoff diffraction formula with $\theta_w^{n}$ denoting the angle made by $n$ th receive antenna with the normal to the RIS at $w$ th element. Using (4) and (5), the effective MIMO channel matrix $\boldsymbol{\tilde{H}}$
entries are written as 
\begin{equation}
\begin{aligned}
[\boldsymbol{\tilde{H}}]_{n,m} &= \sum_{a=0}^{W_x-1} \sum_{b=0}^{W_y-1}\sqrt{\mathbf{PL}_{a,b}^{m}G_{n}}\frac{L_xL_y}{j\lambda D_{a,b}^{n}}F(\theta_w^{n}) \\
&\quad\quad e^{-jk(D_{a,b}^{m}-D_{a,b}^{n})} e^{j\varphi_{a,b}}e^{-(\kappa_{\text{abs}}(f)/2)(D_{a,b}^{n})}.
\end{aligned}
\end{equation}
To model the direct link between the Tx and Rx, the LoS channel model is employed, whose channel coefficient is given by [3]
\begin{equation}
[\boldsymbol{H}]_{n,m} = \frac{\sqrt{G_{n} G_{m}e^{-\kappa_{\text{abs}}(f)(D^{n,m})}} \lambda}{4\pi D^{n,m}} e^{-j k D^{n,m}},
\end{equation}
where $D^{n,m} \triangleq \left\|\mathbf{p}_{n} -\mathbf{p}_{m}\right\|$, respectively.

\section{Problem Formulation and Solution}

To determine the fundamental capacity limit, we make the assumption that both the Tx and Rx possess CSI. Consequently, the MIMO channel capacity in bits per second per Hertz (bps/Hz) is defined as follows:
\begin{equation}
C = \max_{\boldsymbol{R_s}:\text{tr}(\boldsymbol{R_s}) \leq P,\boldsymbol{R_s} \succeq 0} \log_2 \det \left( \boldsymbol{{I}}_{N_r} + \frac{1}{\sigma^2} \boldsymbol{\tilde{H}} \boldsymbol{R_s} \boldsymbol{\tilde{H}}^H \right).
\end{equation}

The capacity of the MIMO channel facilitated by the RIS, presented in (8), is intrinsically connected to the RIS transmission matrix, ${\boldsymbol{\it \Phi}}$. This matrix impacts both the effective channel matrix, $\boldsymbol{\tilde{H}}$, and the resulting optimal transmit covariance matrix, $\boldsymbol{R_s}$. The objective is to maximize the capacity by jointly optimizing ${\boldsymbol{\it \Phi}}$ and $\boldsymbol{R_s}$, under the constraints that dictate the total transmit power and the uni-modular nature of the transmission coefficients. The formulation of this optimization problem is as follows:
\begin{align}
 \text{(P)}\quad \max_{{\boldsymbol{\it \Phi}},\boldsymbol{R_s}} \quad & \log_2 \det \left( \boldsymbol{I}_{N_r} + \frac{1}{\sigma^2} \boldsymbol{\tilde{H}} \boldsymbol{R_s} \boldsymbol{\tilde{H}}^H  \right) \label{eq:P1_obj} \\
\text{s.t.} \quad & {\boldsymbol{\it \Phi}} = \text{diag}{\{\beta_1, \dots, \beta_W\}} \label{eq:P1_const1} \\
& |\beta_w| = 1, \quad w = 1, \dots, W \label{eq:P1_const2} \\
& \text{tr}(\boldsymbol{R_s}) \leq P \label{eq:P1_const3} \\
& \mathbf{\boldsymbol{R_s}} \succeq \boldsymbol{0}.
\label{eq:P1_const4}
\end{align}

The solution to this optimization problem is derived from the analysis presented in [10]. An alternating optimization algorithm is applied to solve (P), subsequent to the transformation of the objective function of (P) in terms of the optimization variables in $\{ \beta_w \}_{w=1}^{W} \cup \{ \boldsymbol{R_s} \}$. It follows solving two subproblems of (P) to optimize the transmit covariance matrix $\boldsymbol{R_s}$ and a single transmission coefficient $\beta_w$ in ${\boldsymbol{\it \Phi}}$ while keeping all other variables constant.


\subsection{Optimization of $\mathbf{R_s}$ With Given $\{\beta_w\}_{w=1}^W$}

Consider the truncated singular value decomposition (SVD) of $\boldsymbol{\tilde{H}}$ to be represented as $\boldsymbol{\tilde{{H}} = \tilde{{A}} \tilde{{\Delta}} \tilde{{B}}^H}$. Here, $\boldsymbol{\tilde{B}}\in\mathbb{C}^{N_t \times E}$, where $E = \text{rank}(\boldsymbol{\tilde{H}}) \leq \min(N_r, N_t)$ signifies the maximum quantity of data streams that can be transmitted over $\boldsymbol{\tilde{H}}$. The optimal $\boldsymbol{R_s}$ is expressed as
\begin{equation}
\boldsymbol{R_s}^* = \boldsymbol{\tilde{{B}}} \text{diag}\{p_1^*, \dots, p_E^*\} \boldsymbol{\tilde{{B}}}^H, \label{eq:Q_opt_given_alpha}
\end{equation}
where $p_i^*$ represents the optimal power allocated to the $i$th data stream adhering to the water-filling strategy: $p_i^* = \max\left({1}/{p_0} - {\sigma^2}/{[\tilde{\mathbf{\Delta}}]^2_{i,i}}, 0\right),$ $i = 1, \dots, E$. The parameter $p_0$ is determined such that the condition $\sum_{i=1}^E p_i^* = P$ is satisfied. Therefore, the channel capacity with given $\{\beta_w\}_{w=1}^{W}$ is 
\begin{equation}
C = \sum_{i=1}^{E} \log_2\left(1 + \frac{\tilde{[\mathbf{\Delta}]}_{i,i}^2 p^*_i}{\sigma^2}\right).
\end{equation}

\subsection{ Optimization of $\beta_w$ With Given $\boldsymbol{R_s}$ and $\{\beta_i,i \neq w\}_{i=1}^{W}$}

Denote $\boldsymbol{R} = [\boldsymbol{r}_1, ..., \boldsymbol{r}_W]$, $\boldsymbol{T} = [\boldsymbol{t}_1, ..., \boldsymbol{t}_W]^H$ with $\boldsymbol{r}_w \in \mathbb{C}^{N_r \times 1}$ and $\boldsymbol{t}_w \in \mathbb{C}^{N_t \times 1},\hspace{1mm} $$\boldsymbol{R_s} = \boldsymbol{U}_{R_s} \boldsymbol{\Lambda}_{R_s} \boldsymbol{U}_{R_s}^ H$ as the eigenvalue decomposition (EVD) of $\boldsymbol{R_s}$, where $\boldsymbol{U}_{R_s} \in \mathbb{C}^{N_t \times N_t}$ and $\boldsymbol{\Lambda}_{R_s} \in \mathbb{C}^{N_t \times N_t}$, $\boldsymbol{T'} = \boldsymbol{T} \boldsymbol{U}_{R_s} \boldsymbol{\Lambda}_{R_s}^{1/2} = [\boldsymbol{t}_1', ..., \boldsymbol{t}_W']^H \in \mathbb{C}^{W \times N_t}$, where $\boldsymbol{t}_w' = \boldsymbol{\Lambda}_{R_s}^{1/2} \boldsymbol{U}_{R_s}^H \boldsymbol{t}_w \in \mathbb{C}^{N_t \times 1}$. The optimal $\beta_w$ is given by:
\begin{equation}
\beta^*_w = \begin{cases} e^{-j\arg\{\Psi_w\}}, & \text{if } \operatorname{tr}(\boldsymbol{X}^{-1}_w \boldsymbol{Y}_w) \neq 0 \\ 1, & \text{otherwise} \end{cases}
\end{equation}
where 
$\boldsymbol{X}_w = \boldsymbol{I}_{N_r} + \frac{1}{\sigma^2} \left( \sum_{i=1,i \neq w}^W \beta_i \boldsymbol{r}_i \boldsymbol{t}_i'^H \right) \left( \sum_{i=1,i \neq w}^W \beta_i \boldsymbol{r}_i \boldsymbol{t}_i'^H \right)^{H} \\
\quad + \frac{1}{\sigma^2} \boldsymbol{r}_w \boldsymbol{t}_w'^H \boldsymbol{t}_w' \boldsymbol{r}_w^H, \quad \forall w \in \mathcal{W}, \label{eq:Aw}$

$\boldsymbol{Y}_w = \frac{1}{\sigma^2} \boldsymbol{r}_w \boldsymbol{t}_w'^H \left(\sum_{i=1,i \neq w}^{W} \boldsymbol{t}_i' \boldsymbol{r}_i^H \beta_i^* \right), \quad \forall w \in \mathcal{W}, \label{eq:Bw}$      

$\Psi_w \in \mathbb{C}$ denotes the sole non-zero eigenvalue of $\boldsymbol{X}^{-1}_w \boldsymbol{Y}_w$ and $\mathcal{W} = \{1, \dots, W\}$.

\subsection{Alternating Optimization}

We begin by generating $L > 1$ sets of $\{\beta_w\}_{w=1}^{W}$, where $|\beta_w| = 1, \forall w$ and the phases of $\beta_w$ follow a uniform distribution within the range of $[0, 2\pi)$. For each set of transmission coefficients, we calculate the optimal transmit covariance matrix $\boldsymbol{R_s}$ using equation (14) and determine the associated channel capacity. The set yielding the maximum capacity is selected as the starting point for the algorithm. The algorithm then progresses by iteratively solving the two subproblems until the convergence is achieved.

\section{Simulation Results}

\begin{figure}
\centering
\includegraphics[width=1\linewidth]{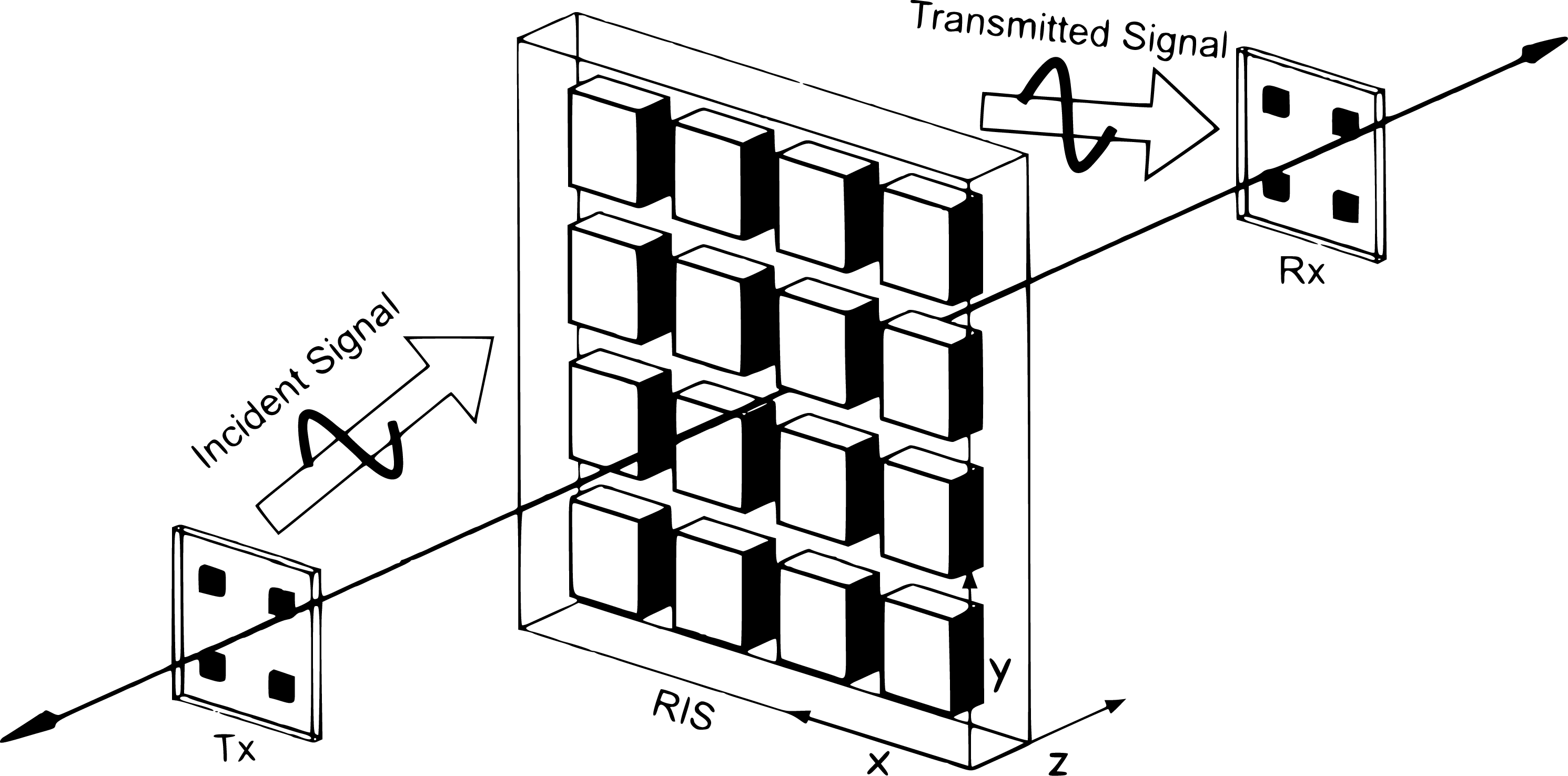}
\caption{Illustration of the simulation setup.}
\label{fig:my_picture}
\end{figure}

To understand the position of the RIS between the Tx and Rx that maximizes the capacity, we examine the setup depicted in Fig. 2. Uniform planar arrays are utilized at both the Tx and Rx. It is assumed that the arrays and the RIS are all in the same plane (the XY-plane), with their centers in alignment. We now fix the positions of Tx and Rx and consider different locations of RIS by varying its z-coordinate to find the maximum capacity corresponding to each location using the algorithm specified in section-\uppercase\expandafter{\romannumeral 3}. Throughout the analysis, we consider the RIS  to be large enough to block all direct paths between the Tx and Rx.

We assess a MIMO configuration with $N_t$ = $N_r$ = 4, i.e., $2 \times 2$-element uniform planar arrays, and a RIS of 1600 elements, i.e., $40 \times 40$  for different inter-antenna spacing in the Tx and Rx. We consider the Tx, Rx, and RIS centers at $(x=0.0122, y=0.0122)$ and fix the Tx and Rx at equal distances from the plane $z = 0$, with the z-coordinates $0.2629$ and $-0.2629$, respectively. We now vary the z-coordinate of the RIS across 50 evenly spaced locations in the range $(-0.1860, 0.1860)$ for each inter-antenna distance of Tx and Rx. Note that all the distances are in meters. We set the Carrier frequency $f$ as 300 GHz, $\kappa_{\text{abs}}(f) = 0.0033 {\rm m}^{-1}$, $G_{m} = G_{n}$ = 20 dBi, $P$ = 10 dBm, $\sigma^2 = -90$ dBm, $L_x$ = $L_y$ = $\lambda$/2, $d_x$ = $d_y$ = $\lambda/8$. For the algorithm, the number of random initializations is established as $L$ = 100, and the convergence threshold, in terms of the relative increase in the objective value, is set to $\gamma = 10^{-5}$. It is important to highlight that the inter-antenna spacing variation for both the Tx and Rx is equivalent. The performance of the RIS-assisted MIMO is compared with LoS MIMO. The maximum capacity of the LoS MIMO is calculated by determining the corresponding optimal transmit covariance matrix $R_s$ using (14).

As depicted in Fig. 8, the maximum achievable capacity of LoS MIMO demonstrates a progressive increase with the augmentation of the inter-antenna spacing at the Tx and Rx. This is because, a greater spacing between the antennas enables improved spatial signal separation and enhances the curvature of the wavefront across the arrays, as discussed in prior studies [3] and [11]. This contributes to increased DoF and multiplexing gain, which can be interpreted from Figs. 4 and 5. The data rate of each stream (as per (15)) depends on the singular value magnitude and power allocated via the water-filling strategy. For the case 2$\lambda$, high singular value variance resulted in no power assigned to the least singular value stream, reducing DoF.
From 2$\lambda$ to 4$\lambda$, the increment in achievable capacity is attributed to an increase in both DoF and multiplexing gain. Once maximum DoF is achieved, the increment is because of the improved multiplexing gain, which can be noted from case 4$\lambda$ to 7$\lambda$ to 10$\lambda$.

The integration of RIS extends the near-field, rendering the curvature of radiated wavefronts more prominent over the arrays. This extension augments singular values (Figs. 4, 6), enhancing DoF (Fig. 5(a), 7(a)). The position corresponding to the maximum capacity depends on the gains achieved through beamforming and multiplexing. For scenarios involving shorter inter-antenna distances, the high variance of singular values (Fig. 3) leads to substantial beamforming gain, positioning the RIS closer to the Tx or Rx for optimal system performance. However, as antenna spacing widens, the dominance of multiplexing gain over beamforming gain due to low variance (Fig. 3) necessitates shifting the optimal position away from the Tx and Rx (Fig. 8). 

\begin{figure}
\centering
\includegraphics[width=0.65\linewidth]{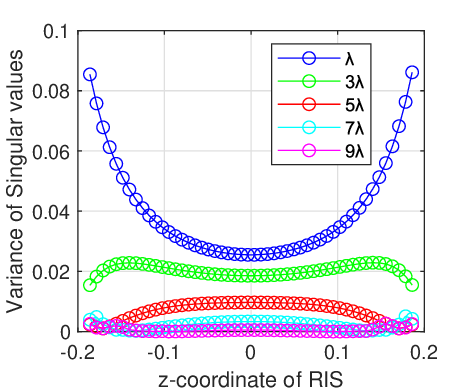}
\caption{Variance of singular values of the channel versus z-coordinate of RIS.}
\label{fig:my_picture}
\end{figure}

\begin{figure*}
    \centering
    \begin{subfigure}{0.23\textwidth}
        \includegraphics[width=\linewidth]{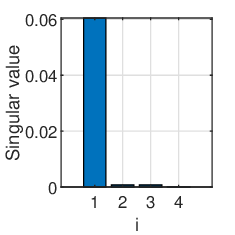}
        \caption{Interantenna spacing of $2\lambda$}
    \end{subfigure}
    \quad
   \begin{subfigure}{0.23\textwidth}
        \includegraphics[width=\linewidth]{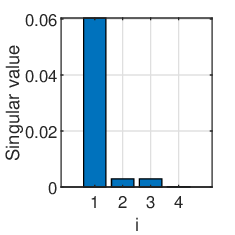}
        \caption{Interantenna spacing of $4\lambda$}
    \end{subfigure}
    \quad
   \begin{subfigure}{0.23\textwidth}
        \includegraphics[width=\linewidth]{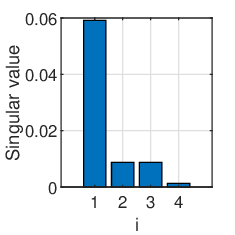}
        \caption{Interantenna spacing of $7\lambda$}
    \end{subfigure}
    \quad
   \begin{subfigure}{0.23\textwidth}
        \includegraphics[width=\linewidth]{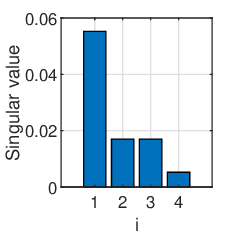}
        \caption{Interantenna spacing of $10\lambda$}
    \end{subfigure}
   
    \caption{LoS MIMO: Singular value versus data stream.}
\end{figure*}

\begin{figure*}
    \centering
    \begin{subfigure}{0.23\textwidth}
        \includegraphics[width=\linewidth]{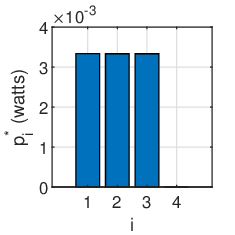}
        \caption{Interantenna spacing of $2\lambda$}
    \end{subfigure}
    \quad
   \begin{subfigure}{0.23\textwidth}
        \includegraphics[width=\linewidth]{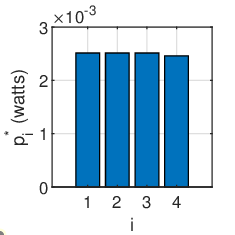}
        \caption{Interantenna spacing of $4\lambda$}
    \end{subfigure}
    \quad
   \begin{subfigure}{0.23\textwidth}
        \includegraphics[width=\linewidth]{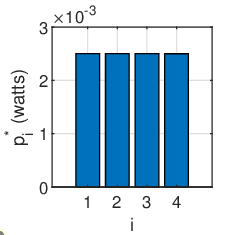}
        \caption{Interantenna spacing of $7\lambda$}
    \end{subfigure}
    \quad
   \begin{subfigure}{0.23\textwidth}
        \includegraphics[width=\linewidth]{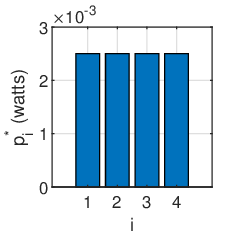}
        \caption{Interantenna spacing of $10\lambda$}
    \end{subfigure}
   
    \caption{LoS MIMO: Optimal power allocated versus data stream.}
\end{figure*}

\begin{figure*}
    \centering
    \begin{subfigure}{0.23\textwidth}
        \includegraphics[width=\linewidth]{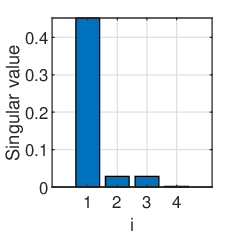}
        \caption{Interantenna spacing of $2\lambda$}
    \end{subfigure}
    \quad
   \begin{subfigure}{0.23\textwidth}
        \includegraphics[width=\linewidth]{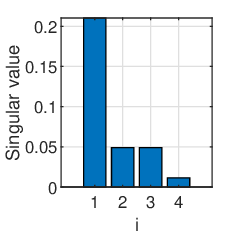}
        \caption{Interantenna spacing of $4\lambda$}
    \end{subfigure}
    \quad
   \begin{subfigure}{0.23\textwidth}
        \includegraphics[width=\linewidth]{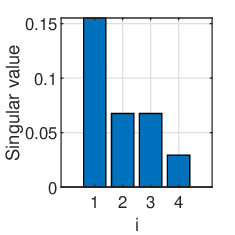}
        \caption{Interantenna spacing of $7\lambda$}
    \end{subfigure}
    \quad
   \begin{subfigure}{0.23\textwidth}
        \includegraphics[width=\linewidth]{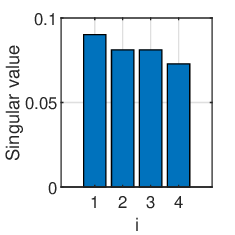}
        \caption{Interantenna spacing of $10\lambda$}
    \end{subfigure}
   
    \caption{RIS aided MIMO: Singular value versus data stream of locations corresponding to maximum capacity.}
\end{figure*}

\begin{figure*}
    \centering
    \begin{subfigure}{0.23\textwidth}
        \includegraphics[width=\linewidth]{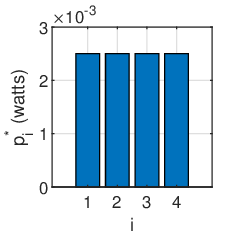}
        \caption{Interantenna spacing of $2\lambda$}
    \end{subfigure}
    \quad
   \begin{subfigure}{0.23\textwidth}
        \includegraphics[width=\linewidth]{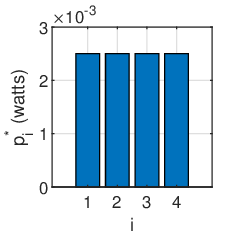}
        \caption{Interantenna spacing of $4\lambda$}
    \end{subfigure}
    \quad
   \begin{subfigure}{0.23\textwidth}
        \includegraphics[width=\linewidth]{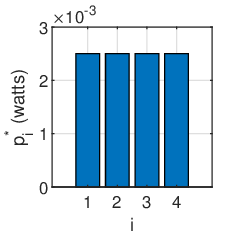}
        \caption{Interantenna spacing of $7\lambda$}
    \end{subfigure}
    \quad
   \begin{subfigure}{0.23\textwidth}
        \includegraphics[width=\linewidth]{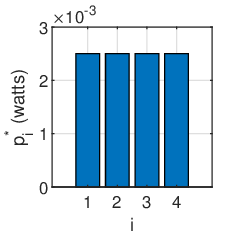}
        \caption{Interantenna spacing of $10\lambda$}
    \end{subfigure}
   
    \caption{RIS aided MIMO: Optimal power allocated versus data stream of locations corresponding to maximum capacity.}
\end{figure*}

\begin{figure*}
    \centering
    \begin{subfigure}{0.3\textwidth}
        \includegraphics[width=\linewidth]{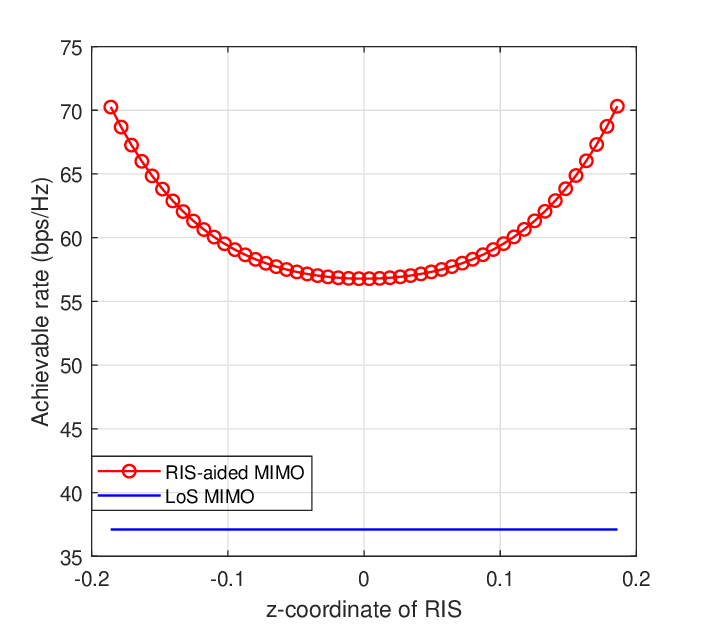}
        \caption{Inter-antenna spacing of $\lambda$}
    \end{subfigure}
    \quad
    \begin{subfigure}{0.3\textwidth}
        \includegraphics[width=\linewidth]{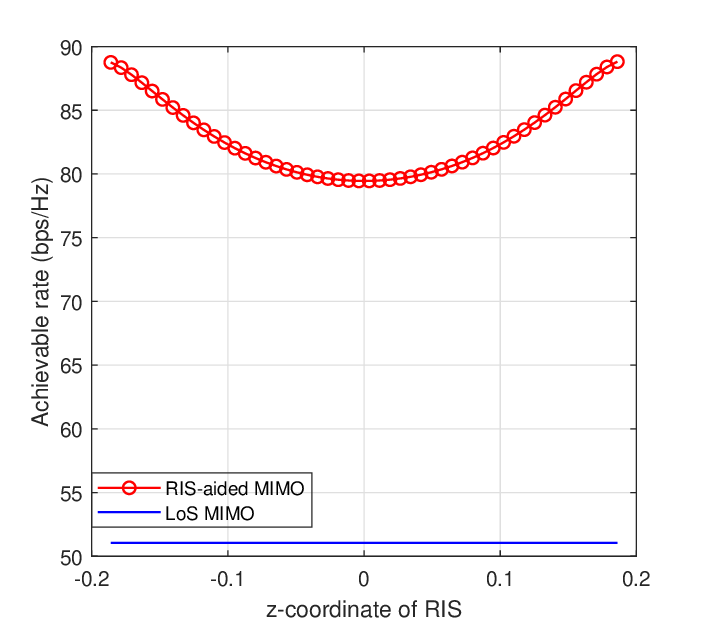}
        \caption{Inter-antenna spacing of 3$\lambda$}
    \end{subfigure}
    \quad
    \begin{subfigure}{0.3\textwidth}
        \includegraphics[width=\linewidth]{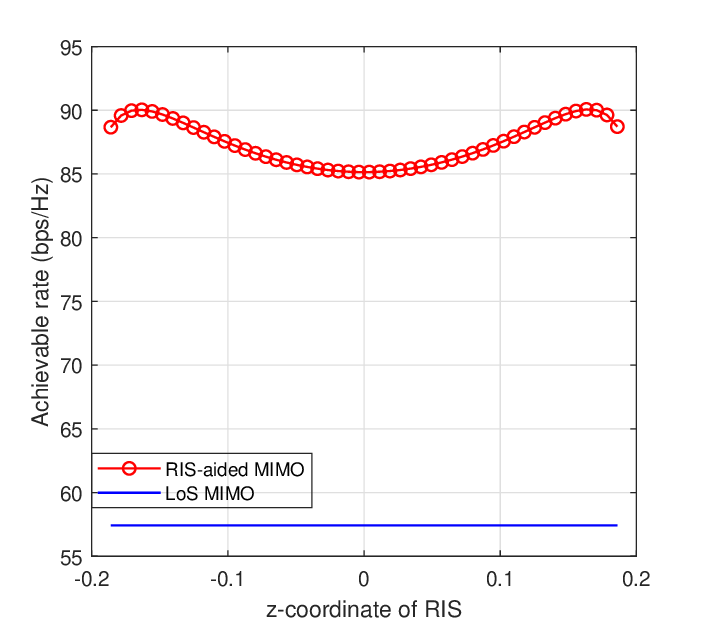}
        \caption{Inter-antenna spacing of 4$\lambda$}
    \end{subfigure}
    \quad
    \begin{subfigure}{0.3\textwidth}
        \includegraphics[width=\linewidth]{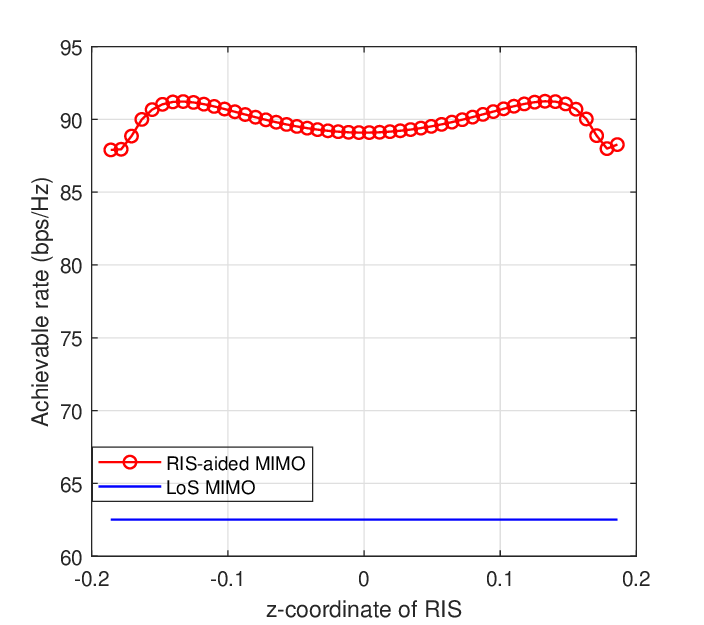}
        \caption{Inter-antenna spacing of $5\lambda$}
    \end{subfigure}
    \quad
    \begin{subfigure}{0.3\textwidth}
        \includegraphics[width=\linewidth]{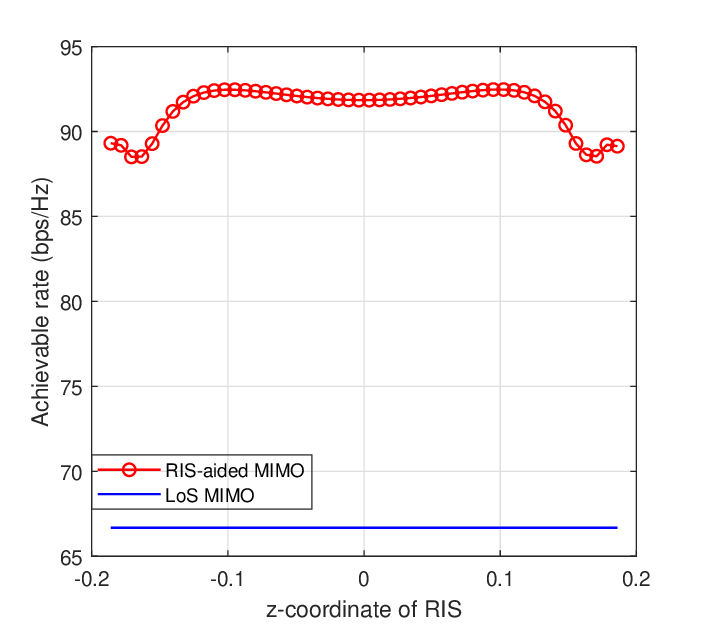}
        \caption{Inter-antenna spacing of 6$\lambda$}
    \end{subfigure}
    \quad
    \begin{subfigure}{0.3\textwidth}
        \includegraphics[width=\linewidth]{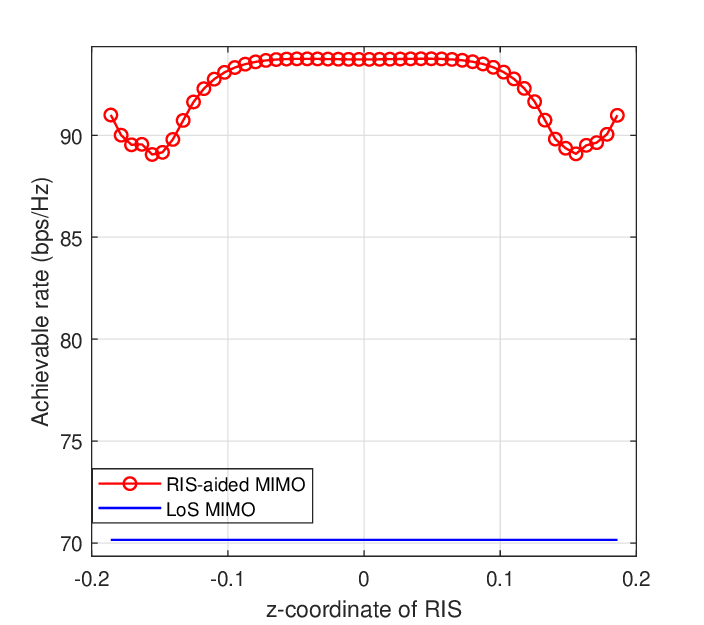}
        \caption{Inter-antenna spacing of 7$\lambda$}
    \end{subfigure}
    \quad
      \begin{subfigure}{0.3\textwidth}
        \includegraphics[width=\linewidth]{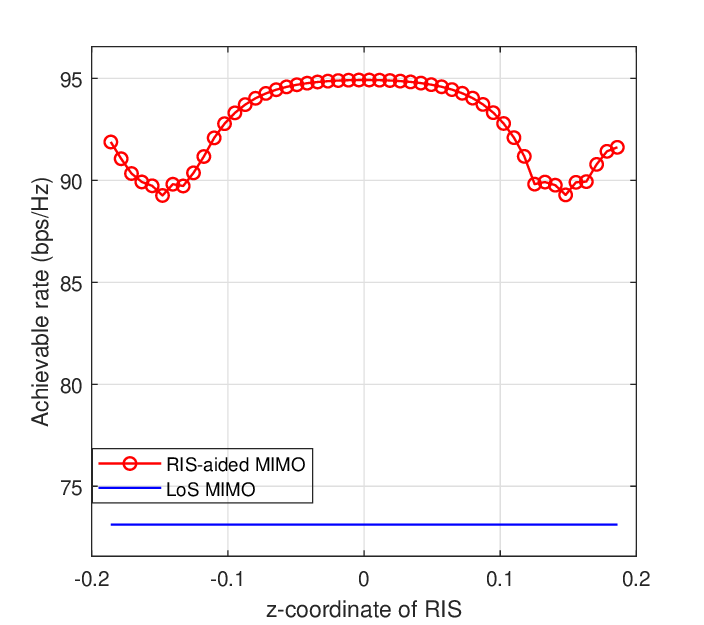}
        \caption{Inter-antenna spacing of $8\lambda$}
    \end{subfigure}
    \quad
    \begin{subfigure}{0.3\textwidth}
        \includegraphics[width=\linewidth]{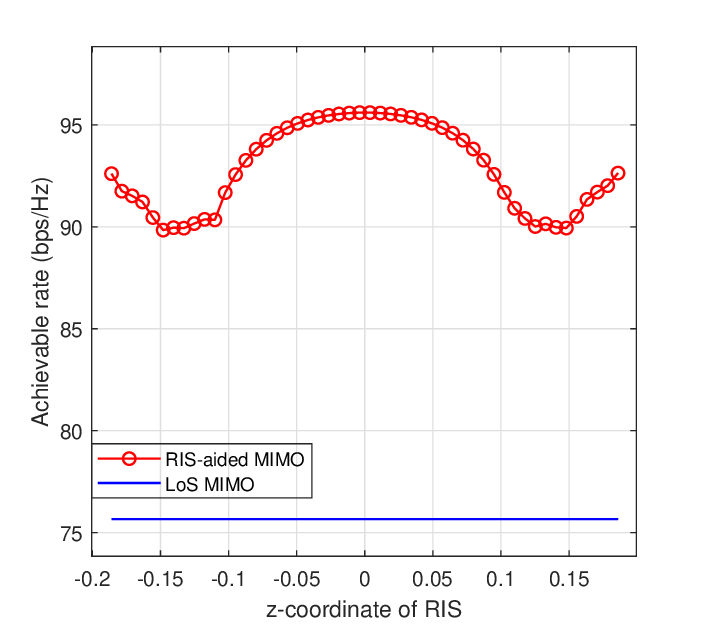}
        \caption{Inter-antenna spacing of 9$\lambda$}
    \end{subfigure}
    \quad
    \begin{subfigure}{0.3\textwidth}
        \includegraphics[width=\linewidth]{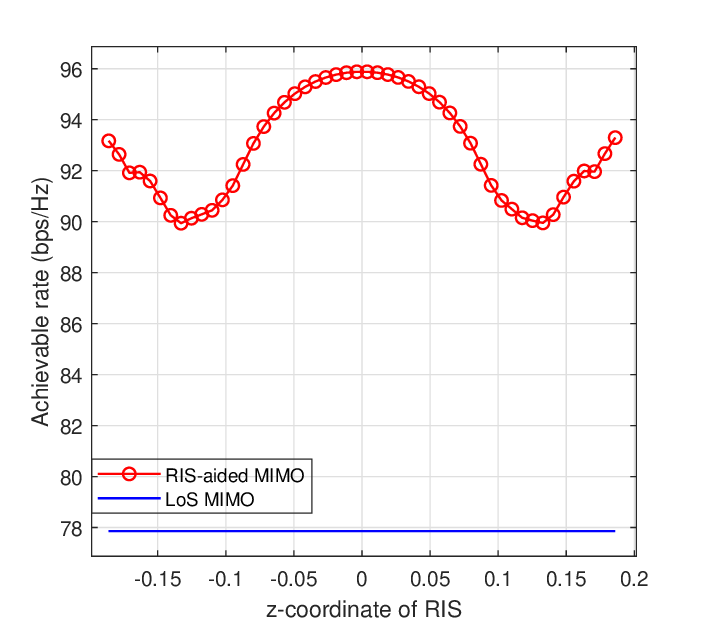}
        \caption{Inter-antenna spacing of 10$\lambda$}
    \end{subfigure}
    \caption{Achievable rate versus z-coordinate of RIS.}
\end{figure*}

\section{Conclusion }

We have analyzed the potential of RIS to extend the near-field region and enhance the spatial multiplexing capabilities of LoS MIMO systems. This RIS functionality can significantly strengthen the wireless backhaul performance at Terahertz (THz) frequencies. We have studied the optimal placement of the RIS in the near field by considering various locations, and through an alternating optimization algorithm, determined the maximum capacity associated with each position. Our findings reveal that the optimal position for the RIS in the near field does not necessarily coincide with the proximity of either the Tx or Rx, as in the far field. Instead, it hinges on the inter-antenna spacing of the Tx and Rx. This observed trend is attributed to the capacity improvement due to multiplexing being more significant than beamforming.


\begin{thebibliography}{00}

\bibitem{}H. -J. Song and T. Nagatsuma, "Present and Future of Terahertz Communications," in {\em IEEE Transactions on Terahertz Science and Technology}, vol. 1, no. 1, pp. 256-263, Sept. 2011, doi: 10.1109/TTHZ.2011.2159552.

\bibitem{}R. Piesiewicz et al., "Short-Range Ultra-Broadband Terahertz Communications: Concepts and Perspectives," in {\em IEEE Antennas and Propagation Magazine}, vol. 49, no. 6, pp. 24-39, Dec. 2007, doi: 10.1109/MAP.2007.4455844.

\bibitem{} H. Do, N. Lee and A. Lozano, "Line-of-Sight MIMO via Intelligent Reflecting Surface," in {\em IEEE Transactions on Wireless Communications}, vol. 22, no. 6, pp. 4215-4231, June 2023, doi: 10.1109/TWC.2022.3224035.

\bibitem{}R. Singh and D. Sicker, "Reliable THz Communications for Outdoor based Applications- Use Cases and Methods," {\em 2020 IEEE 17th Annual Consumer Communications \& Networking Conference (CCNC)}, Las Vegas, NV, USA, 2020, pp. 1-4, doi: 10.1109/CCNC46108.2020.9045670.

\bibitem{}E. Basar, M. Di Renzo, J. De Rosny, M. Debbah, M. -S. Alouini and R. Zhang, "Wireless Communications Through Reconfigurable Intelligent Surfaces," in {\em IEEE Access}, vol. 7, pp. 116753-116773, 2019, doi: 10.1109/ACCESS.2019.2935192.

\bibitem{} J. Dang, Z. Zhang and L. Wu, "Joint beamforming for intelligent reflecting surface aided wireless communication using statistical CSI," in {\em China Communications}, vol. 17, no. 8, pp. 147-157, Aug. 2020, doi: 10.23919/JCC.2020.08.012.

\bibitem{} Dang, T. M., Quoc, T. V., Van Tuan, L., \& Nguyen, H. Q. (2021). A geometry-based stochastic channel model and its application for intelligent reflecting surface assisted wireless communication. {\em IET Communications}, 15(13), 1693-1702.


\bibitem{} K. Dovelos, S. D. Assimonis, H. Quoc Ngo, B. Bellalta and M. Matthaiou, "Intelligent Reflecting Surfaces at Terahertz Bands: Channel Modeling and Analysis," {\em 2021 IEEE International Conference on Communications Workshops (ICC Workshops)}, Montreal, QC, Canada, 2021, pp. 1-6, doi: 10.1109/ICCWorkshops50388.2021.9473890.


\bibitem{} J. Xu, Y. Liu, X. Mu and O. A. Dobre, "STAR-RISs: Simultaneous Transmitting and Reflecting Reconfigurable Intelligent Surfaces," in {\em IEEE Communications Letters}, vol. 25, no. 9, pp. 3134-3138, Sept. 2021, doi: 10.1109/LCOMM.2021.3082214.

\bibitem{} S. Zhang and R. Zhang, "On the Capacity of Intelligent Reflecting Surface Aided MIMO Communication," {\em 2020 IEEE International Symposium on Information Theory (ISIT)}, Los Angeles, CA, USA, 2020, pp. 2977-2982, doi: 10.1109/ISIT44484.2020.9174375.

\bibitem{} Jeng-Shiann Jiang and M. A. Ingram, "Spherical-wave model for short-range MIMO," in {\em IEEE Transactions on Communications}, vol. 53, no. 9, pp. 1534-1541, Sept. 2005, doi: 10.1109/TCOMM.2005.852842.




\end{thebibliography}
\end{document}